\documentclass[useAMS,usenatbib]{mn2e}

\usepackage{graphicx}

\title[Stellar activity as noise in exoplanet detection]{Stellar activity as noise in exoplanet detection I. Methods and application to solar-like stars and activity cycles}
\author[H. Korhonen et al.]{H. Korhonen$^{1,2,3}$\thanks{E-mail: heidi.h.korhonen@utu.fi}, 
J. M. Andersen$^{4,3}$, N. Piskunov$^{5}$, T. Hackman $^{1,6}$, D. Juncher$^{2,3}$, 
\newauthor
S. P. J{\"a}rvinen$^{7}$, and U. G. J{\o}rgensen$^{2,3}$\\
$^{1}$Finnish Centre for Astronomy with ESO (FINCA), University of Turku, 
V{\"a}is{\"a}l{\"a}ntie 20, FI-21500 Piikki{\"o}, Finland\\
$^{2}$Niels Bohr Institute, University of Copenhagen, Juliane Maries Vej 30, 
DK-2100 Copenhagen, Denmark\\
$^{3}$Centre for Star and Planet Formation, Natural History Museum of Denmark, 
University of Copenhagen, {\O}ster Voldgade 5-7, \\ 
DK-1350, Copenhagen, Denmark\\
$^{4}$Department of Astronomy, Boston University, 725 Commonwealth Avenue, 
Boston, MA 02215, USA\\
$^{5}$Department of Physics and Astronomy, Uppsala University, Box 515, S-75120
Uppsala, Sweden\\
$^{6}$Department of Physics, PO Box 64, FI-00014 University of Helsinki, Finland\\
$^{7}$Leibniz-Institut f{\"u}r Astrophysik Potsdam (AIP), An der Sternwarte 16, D-14482 Potsdam, Germany
}
\begin{document}

\date{Accepted 1988 December 15. Received 1988 December 14; in original form 1988 October 11}

\pagerange{\pageref{firstpage}--\pageref{lastpage}} \pubyear{2002}

\maketitle

\label{firstpage}

\begin{abstract}
The detection of exoplanets using any method is prone to confusion due to the intrinsic variability of the host star. We investigate the effect of cool starspots on the detectability of the exoplanets around solar-like stars using the radial velocity method. For investigating this activity-caused "jitter" we calculate synthetic spectra using radiative transfer, known stellar atomic and molecular lines, different surface spot configurations, and an added planetary signal. Here, the methods are described in detail, tested and compared to previously published studies. The methods are also applied to investigate the activity jitter in old and young solar-like stars, and over a solar-like activity cycles. We find that the mean full jitter amplitude obtained from the spot surfaces mimicking the solar activity varies during the cycle approximately between 1~m/s and 9~m/s. With a realistic observing frequency a Neptune mass planet on a one year orbit can be reliably recovered. On the other hand, the recovery of an Earth mass planet on a similar orbit is not feasible with high significance. The methods developed in this study have a great potential for doing statistical studies of planet detectability, and also for investigating the effect of stellar activity on recovered planetary parameters.
\end{abstract}

\begin{keywords}
planets and satellites: detection -- stars: activity -- stars: rotation -- stars: solar-type -- (stars:) starspots
\end{keywords}

\section{Introduction}

The search for exoplanets has traditionally concentrated on stars with very little intrinsic activity. Studies have shown that the known exoplanet host stars exhibit very low levels of magnetic activity (e.g., \citealt{Jenkins06, Martinez10}). Still, as the Kepler satellite has shown, many solar-like stars are more active than our Sun (e.g., Basri et al. 2013), and therefore show significant levels of activity, which can affect the planet detection. The spectral line-profile variations caused by starspots have been confused with a radial velocity signal originating from exoplanets (e.g., \citealt{Queloz01, Huerta08}) and the sudden brightenings caused by stellar flares can mimic microlensing events from a planet sized body (e.g., \citealt{Bennett12}). Magnetic activity, and the phenomena related to it, are an integral part of stars with spectral types ranging from mid-F to M, and as these are the stars most exoplanet searches concentrate on, it is crucial to understand the effects activity sets on exoplanet detection and parameter determination.

The exoplanet detection method most prone to confusion from stellar activity is the radial velocity search. Already, \cite{SaarDonahue97} showed that cool spots on the stellar surface cause spectral line-profile variations that can be confused with the radial velocity variations from planets. In addition, they derived an analytical formula to represent the relation of this radial velocity jitter to the fraction of stellar surface covered by starspots (starspot filling factor). Other similar investigations have been carried out, see, e.g., \cite{Desort07}, \cite{Reiners10}, \cite{Dumusque11b}, \cite{Boisse11}, and \cite{Barnes11}. Also, \cite{Boisse12} provided a freely available tool, SOAP (Spot Oscillation And Planet), for the community to investigate the effects of starspots on radial velocity measurements and photometric observations. The code was expanded and improved by \cite{Oshagh13} to include planetary transits on a spotted star, and a new modified version, SOAP 2.0, was recently published by \cite{Dumusque14}.

Most planet searches have concentrated on solar-like stars. Therefore, there have also been relatively many investigations on the effect of solar activity on exoplanet detection (see, e.g., \citealt{Lagrange10}; \citealt{Meunier10}). \cite{Meunier13} used solar activity as a template to study the effect of spots and plages on detectability of Earth-mass exoplanets in the habitable zone of their host star. They conclude that especially the contribution from the plages would prevent the detection of Earth around the Sun, even with forthcoming high precision instruments. These investigations have concentrated on plages and spots, but these are not the only error sources. Granulation and stellar oscillations also cause noise at m/s level (see, e.g., \citealt{Dumusque11a}), which will hinder the detection of small sized planets, and planets on wide orbits. Still, the timescale for these noise sources is much shorter than the variations caused by long-period planets, therefore they can be averaged out using long exposures and/or observing frequently. A new potential noise source, gravitational redshift, was identified recently by \cite{Cegla12}, but its magnitude is estimated to be only few cm/s.

Photometric observations have been used to estimate the activity-induced jitter in radial velocity measurements (e.g., \citealt{Lanza11}; \citealt{Aigrain12}). In a recent work based on GALEX ultraviolet measurements and Kepler light-curves \cite{Cegla14} investigate how well the radial velocity jitter can be estimated based on photometry alone. They conclude that for magnetically quiet stars one can use photometric measurements as a proxy for radial velocity variability.

\cite{MaGe12} developed a technique to use the radial velocity method for detecting planets even when the host star shows significant activity. Their method relies on the wavelength dependence of the spot-caused jitter as opposed to the planetary signature which is not wavelength dependent. \citet{Moulds13} have also shown that for active stars it is possible to remove some of the spot signature from the line-profiles and still be able to recover Jupiter sized close-in planets.

In this work we develop methods that use radiative transfer to calculate spectral-line profiles from a spotted stellar surface. The stellar spot configurations are either based on spot sizes and numbers, or filling factors. It is also possible to introduce active longitudes and latitudes. In addition, a planetary signal can be added to the spectra calculated from the spotted surface. This approach allows for a statistical investigation of exoplanet detection and enables obtaining information on the errors the stellar activity causes on the determined planetary parameters. In the current paper, Paper I, we describe the methods, test them and compare the results to some of the previously published studies. The methods are also applied to study jitter around solar-like stars, including a solar-like activity cycle. In the second part, Paper II, we will apply the methods to M dwarfs and investigate the reliability of recovery of planetary parameters in the presence of stellar activity (\citealt{Andersen14}).

\section{Methods}
We create spot patterns on a simulated stellar surface using our SPOTSS code (see next Section). Synthetic spectral line profiles at different rotational phases of the star are calculated based on the created surface distributions. The spectra based on the given spot configuration and local line profile grids are calculated using the code DIRECT7, which is written by N.E. Piskunov and includes modifications introduced by T.Hackman. This code uses the same routines as INVERS7 (\citealt{Piskunov90}; \citealt{Hackman01}). Radial velocity (jitter) measurements are obtained by cross-correlation of the calculated line profiles with either one of the generated profiles or a template profile, which is obtained from a spectrum with the temperature of the unspotted stellar surface. Exoplanet radial velocity is introduced into the spectra and recovered by a new cross-correlation. These steps after the spot surface creation are all done in a code called DEEMA (Detection of Exoplanets under the Effect of Magnetic Activity). Details of all the steps are discussed in the following.

\subsection{Generating spots}

We developed our code, SPOTSS, to generate spot (temperature) patterns on a simulated stellar surface. The code can generate random spots across the entire surface, and has also the option to define certain `active' regions in latitude and/or longitude. The code creates a matrix of temperatures which represents the entire stellar surface, corresponding to $N$ points of latitude and $2\times N$ points of longitude. Using a $60\times 120$ grid yields a latitude resolution of 3 degrees/pixel and a longitude resolution of 3 degrees/pixel at the equator (note that the actual longitude resolution increases near the poles since the grid is a flat square that represents a spherical surface). The size of the matrix can be changed in order to increase or decrease the spatial resolution, but at much higher sizes the processing time is very long compared to the relatively small gains in precision.

The code takes the following (user defined) input parameters:

\begin{enumerate}
\item Stellar temperature, $Tp$: the photospheric temperature of the unspotted stellar surface, before spots are added. Every value in the temperature matrix is originally set to this value.
\item Spot temperature, $Ts$: the temperature of the umbral regions of the spots. This is considered the spot temperature, although each spot also has a penumbral region with a temperature defined as the mid point between the spot temperature and the photospheric temperature.
\item Spot radius, $rs$: `average' radius of the spots. The radius of each spot is randomly altered starting from this value to create a lognormal distribution of spot sizes around this size. Lognormal distribution is chosen because the sunspot size distribution is known to follow it (e.g., \citealt{Bogdan88}; \citealt{Baumann05}). Spots are approximately circular, and we add penumbral regions with umbral to penumbral radii ratios of 1:2, creating an umbral to penumbral area of 1:3, following \cite{Solanki03}.
\item Number of spots, ns (or filling factor, depending on which version of the code is used): spots are placed on the stellar surface randomly until the number of spots is reached or the desired filling factor is achieved. For certain purposes it was useful to investigate the effect of one large spot, so ns was set to 1. (Note when ns = 1 the exact input value of rs is used, since it is not necessary to alter this to obtain a certain size distribution of spots. An exact latitude and longitude for the center of the spot can also be specified.)
\item Longitude range: defaults to the full range of longitude: 0 -- 360 degrees, but active longitude ranges can be defined meaning spots will only be placed within those ranges.
\item Latitude range: defaults to the full range of latitude, -90 -- 90 degrees, though active latitudes can also be defined. Then spots will only be placed within those ranges. This can be combined with defining an active longitude range to create a small "active region" on the stellar surface. Active regions such as this have been observed in some Doppler Images of active stars.
\end{enumerate}

\begin{table*}[h]
  \centering
    \caption{Sources of data for the molecular line opacities used in SYNTHE for the M dwarf spectral grid.}
  \begin{tabular}{@{}ccccc@{}}
\hline
Molecule  & Transitions &Wavelength range [{\AA}] & \#Lines & Source\\
\hline
CH & A-X, B-X, C-X & 2600 -- 17,000  & 71,600 & R. Kurucz \\
CN & A-X, B-X & 2000 -- 1,000,000 & 1,645,000 & R. Kurucz\\ 
CO & X-X, A-X & 1100 -- 100,000 & 555,000 & R. Kurucz\\
OH & X-X, A-X & 2000 -- 1,000,000 & 82,000 & R. Kurucz\\
H$_2$O & X-X & 4100 -- 1,000,000 & 65,900,000 & R. Kurucz\\
SiO & X-X, A-X, E-X & 1400 -- 1,000,000 & 1,830,000 & R. Kurucz\\
TiO & A-X, B-X, C-X, E-X, c-a, b-a, b-d, f-a & 4100-1,000,000 & 33,000,000 & D. W. Schwenke\\
\hline
\end{tabular}
\label{molecules}
\end{table*}

\subsection{Calculating spectral line-profiles}

For calculating synthetic spectra we used two different atmospheric models: one for the solar-like stars and one for the M dwarfs. For both cases 17 limb angles (0.01, 0.025, 0.050, 0.075, 0.1, 0.125, 0.15, 0.2, 0.25, 0.3, 0.4, 0.5, 0.6, 0.7, 0.8, 0.9, 1.0) were used. For solar-like stars a grid of local line profiles was created using the SPECTRUM spectral synthesis code \citep{SPECTRUM} and Kurucz model atmospheres \citep{kurucz93}. The grid includes temperatures 4000 -- 6000~K with a temperature step of 250~K. The line lists luke.lst and luke.nir.lst (included in the SPECTRUM package), which include atomic lines and some molecular species, are used in the calculations. For the photometry we use the same code and models but for a sparser wavelength grid ranging from 3600~{\AA} to 7350~{\AA} and step size of 50~{\AA}. 

For the M dwarf temperatures we used the program SYNTHE (\citealt{synthe1}; \citealt{synthe2}) to compute the synthetic spectra. The spectra are based on a subset of the 2008 grid of MARCS stellar model atmospheres (\citealt{gustafsson}) with solar metallicities. They cover a temperature range of $T$ = 2500--4000~K, a surface gravity range of log($g$)= 4.5--5.5 and a micro-turbulence range of $\xi _t$ = 0.0--2.0~m/s. For the atomic line opacities we used data from the on-line database VALD-2 \cite{vald}, and for the molecular line opacities we used the species presented in Table \ref{molecules}. The calculated spectra cover a wavelength range of 3000--9200~{\AA} and have a resolution of $\Delta \lambda / \lambda$ = 500,000. 

The spectra based on the given spot configuration and local line profile grids are calculated using the code DIRECT7. Before running DIRECT7 the local line profiles are convolved with a Gaussian instrumental profile and a radial-tangential macroturbulence (here set to 2~km/s). Furthermore, local fluxes for the B- and/or V-magnitudes are calculated using the transmission functions for these wavelength passbands.

For each rotation phase the visible stellar hemisphere is simulated in DIRECT7 and the spectrum and B- and V-passband fluxes  are calculated by integration of the line and continuum intensities over this visible stellar hemisphere. The projection effects, limb darkening and visibility of the spots are thus naturally directly taken into account. User defined $v\sin i$ and inclination values are applied to the spectral line profiles. The  $v\sin i$ can be given in two ways: either the user can supply the desired value, or the DEEMA code can estimate the value based on the stellar mass, rotation period and inclination. For the estimation models by \cite{Baraffe98} are used. From now on the term 'spectrum' refers to the synthetic spectrum that has been calculated by integration over the full visible stellar disc.

The spectra can be calculated at any given rotational phase. Here, a scheme where the length of the observing run and stellar rotation period are given in days, and the user supplied number of phases is evenly distributed over the observing run and the stellar rotational phases are calculated based on this information. The stellar rotational phases are used as an input for the line profile and B- and/or V-magnitude calculations. 

\begin{figure*}
  \includegraphics[width=16cm]{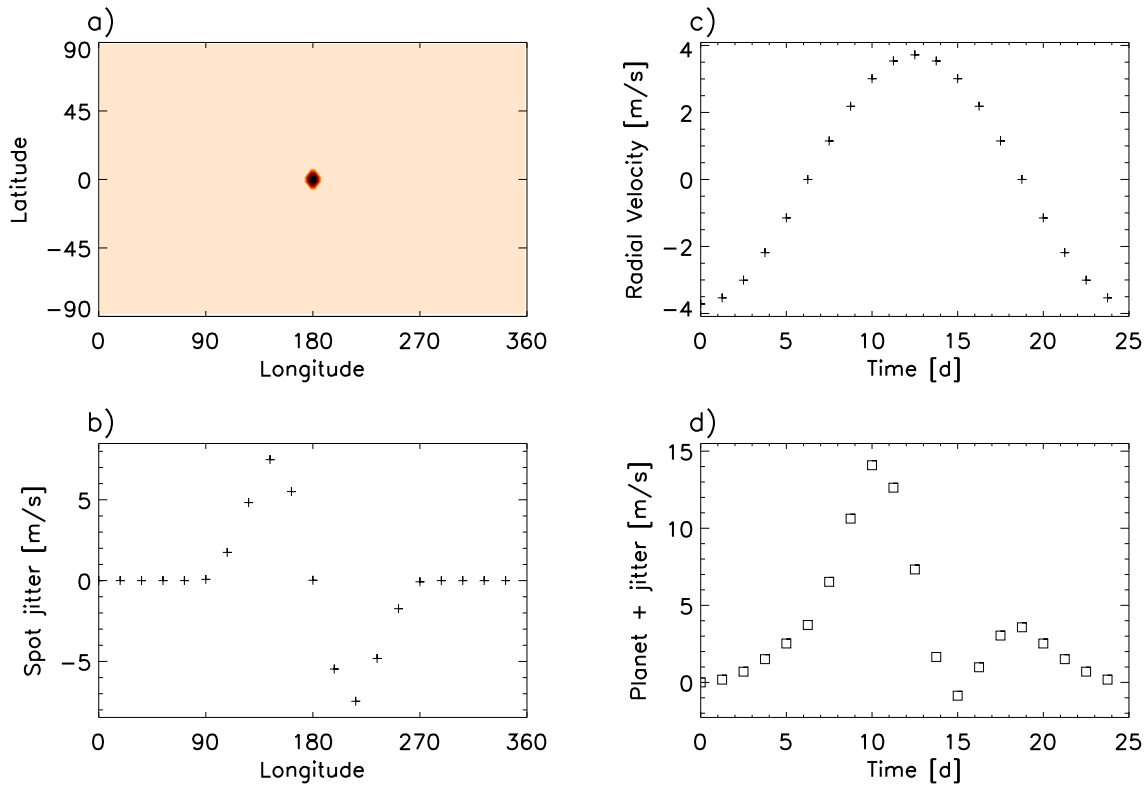}
  \caption{An example of stellar surface spot configuration and the resulting radial velocity jitter from a spot and a planet. a) Stellar surface configuration used for calculating the spectra. The x-axis gives the longitude in degrees and y-axis the latitude, also in degrees. The unspotted surface has a temperature of 5800~K and the spot temperature is 4000~K in the umbra and 4900~K in the penumbra. The radius of the spot is $\sim 5^{\circ}$. b) The resulting jitter curve from the five degree spot. The x-axis is the longitude in degrees and y-axis the measured radial velocity jitter in m/s. The jitter values are calculated at 20 evenly spaced rotational phases from the wavelength region 5952--5998~{\AA}. The error of the jitter measurement is smaller than the symbol size. c) Radial velocity curve of a Neptune mass planet on a 25 day circular orbit around one solar mass star. The x-axis gives the time in days and y-axis the radial velocity in m/s. d) The radial velocity curve from the five degree spot and Neptune sized planet. Owing to the cross-correlation scheme used here (see Section \ref{section_cc}) the first jitter measurement is always shifted to zero. This explains the different absolute values for the input and calculated radial velocities. In the plot x-axis has the time in days and y-axis the measured radial velocity in m/s.}
  \label{Jitter_example}
\end{figure*}

\subsection{Cross-correlation}
\label{section_cc}

For obtaining the jitter values induced by spots, spectra calculated for different observational phases are cross-correlated. The cross-correlation can be carried out against different templates: spectra without spots obtained using unspotted surface temperature, spectra without spots obtained using the mean temperature of the stellar surface, or using one of the spectra with spots as the template. 

If one of the spectra created with surface spots is used as the template, then all the spectra are cross-correlated against each other. This means that there are N-1 jitter curves, where N is the number of rotational phases. All the obtained jitter curves are normalised in such a way that the measurement at the first phase is set to zero, and the mean and the standard deviation are calculated for each phase. This provides a mean jitter curve and an estimate of the measurement error. The jitter curves are virtually identical and the error estimates small, as should be when no noise is added to the spectra. This is the method that is used in all the calculations presented in this work, and owing to this scheme the cross-correlation result is always zero for the first longitude.

In cross-correlation the accuracy is increased by improving the sampling of the cross-correlation curve using linear interpolation and fitting a polynomial to the curve. The maximum of the polynomial is calculated and the corresponding shift used as the jitter value. An example of a stellar surface and resulting jitter curve are shown in Fig.~\ref{Jitter_example}a and \ref{Jitter_example}b, respectively. The spot is a 5$^{\circ}$ radius equatorial spot with umbral temperature of 4000~K, penumbral temperature of 4900~K and unspotted surface temperature of 5800~K. The stellar rotation period has been set to 25 days and the spectra have been calculated at 20 evenly spaced phases using the wavelength region 5952--5998~{\AA}. The inclination is set to 90$^{\circ}$ (equator-on) and $v\sin i$ is estimated from the period and stellar radius to be 2.1 km/s. The resulting jitter curve has a full amplitude of 14.95~m/s.

\begin{figure}
  \includegraphics[width=8cm]{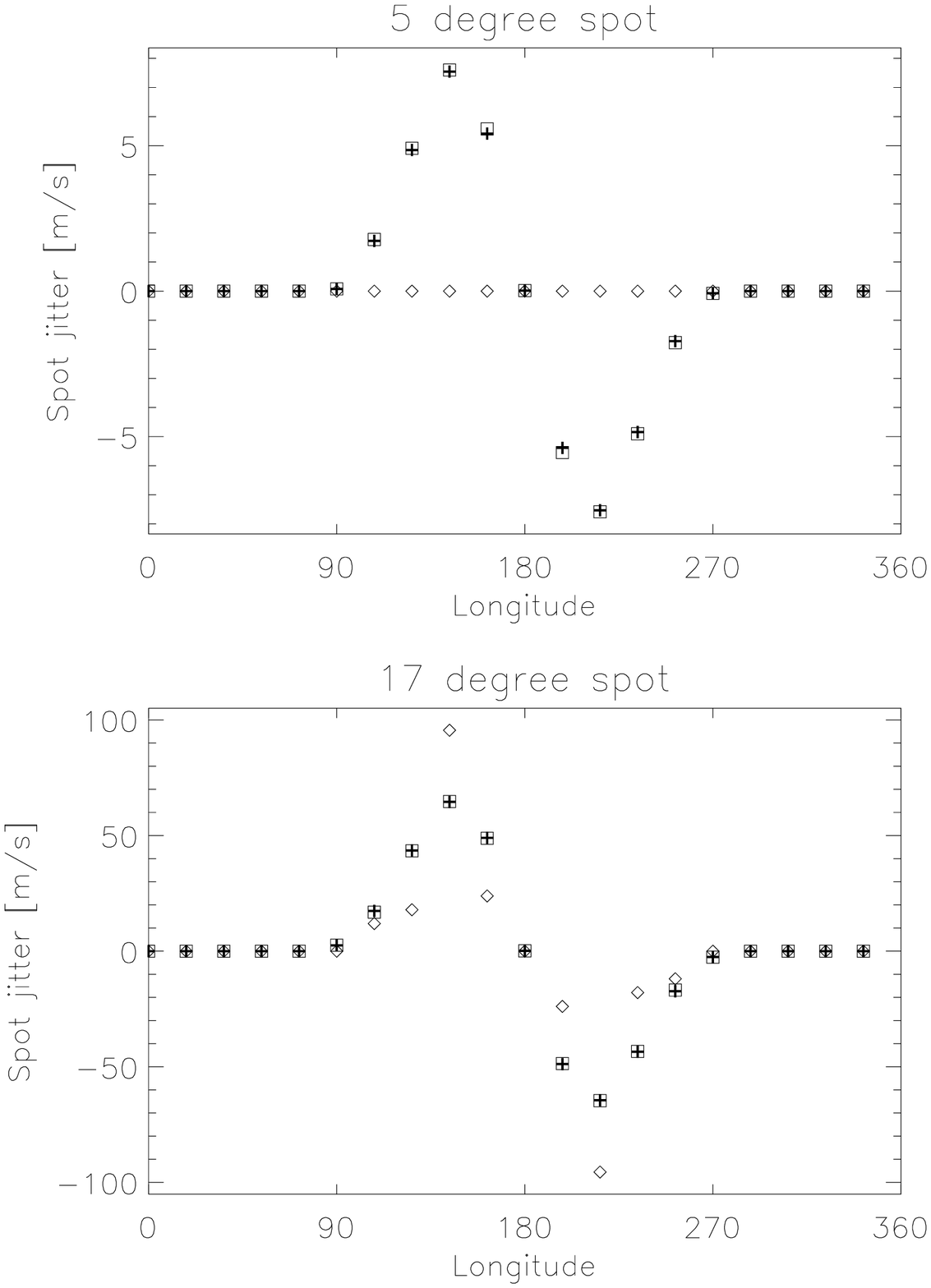}
  \caption{An example of jitter curves obtained using different methods for determining the maximum of the cross-correlation function: polynomial fit (plus-signs), Gaussian function (squares) and taking the maximum of the cross-correlation function (diamonds). The results are shown for two different equatorial spots, one with radius of 5 degrees (top) and one of 17 degrees (bottom). }
  \label{cc_max}
\end{figure}

Tests were also carried out using a Gaussian function instead of a second degree polynomial for determining the maximum of the cross-correlation function. Figure~\ref{cc_max} shows the results for three different methods: polynomial (plus-signs), Gaussian function (squares) and simply taking the maximum of the cross-correlation function (diamonds). The results are shown for two different equatorial spots, one with radius of 5 degrees (top) and one of 17 degrees (bottom). As expected, simply taking the maximum of the cross-correlation function gives zero when the shift is small, and with larger shifts it at times gives exaggeratedly large values. On the other hand, both polynomial and Gaussian fits give very similar results. Based on calculations using 100 different spot configurations with spot filling factor of 0.02\%, it can be seen that the polynomial fit always gives somewhat smaller value than the Gaussian fit. The full jitter amplitude obtained from the polynomial fit is $2.85\pm 0.03$~\% smaller than the value obtained using the Gaussian fit. The tendency for smaller jitter values with the polynomial fit is also seen in Fig.~\ref{cc_max}, but for both the 5 degree and 17 degree radius spots the difference between the methods is less than 1\%. 

\begin{figure}
  \includegraphics[width=8cm]{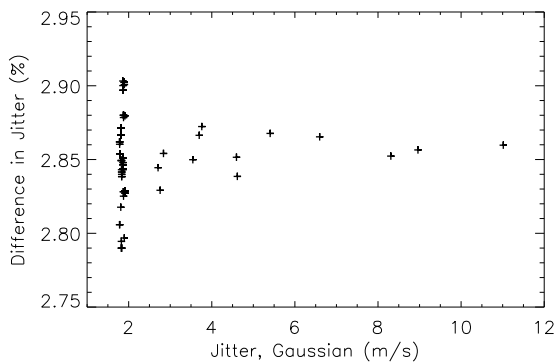}
  \caption{The procentual difference between the full jitter amplitude obtained using polynomial and Gaussian fits plotted against the full jitter amplitude from the Gaussian fit.}
  \label{cc_difference}
\end{figure}

There does not seem to be any correlation between the full jitter amplitude and the performance of the Gaussian and polynomial fits. Figure~\ref{cc_difference} shows the procentual difference between the full jitter amplitude obtained using polynomial and Gaussian fits and plotted against the full jitter amplitude obtained from the Gaussian fit. The larger jitter does not result in larger (or smaller) procentual difference between the methods. On the whole, both Gaussian and polynomial fits give very similar results. In the following polynomial fits are used.

DIRECT7 uses an evenly spaced wavelength grid for calculating the spectra. Our tests show that over such a small wavelength range (50{\AA}) the results are not significantly affected even if a logarithmic wavelength scale is not used in the cross-correlation. Therefore, we use the DIRECT7 output directly in the cross-correlation.

\subsection{Introducing a planet to the spectra}

We generate RV curves resulting from orbiting planets using Kepler's Third Law. The elliptical case of Kepler's equation is solved following the formalism by \cite{Mikkola87}. The three optional variables ($\gamma$, the systemic velocity, or arbitrary instrumental offset; $\dot\gamma$, the systemic acceleration, due to systematics in the data or an additional body in the system with a much longer period; and $t_{0}$, an arbitrary zero point for the slope) were left out. These parameters are used when fitting RV curves, but they are not necessary when simply generating a RV curve. This leaves the equation 
\begin{equation}
RV(t) = K[\cos\theta (t)+ \omega_{*} + e\cos\omega_{*}], 
\end{equation}
where K is the radial velocity semiamplitude and is given in m/s by 
\begin{equation}
K = (\frac{2\pi G}{P(M_{P}+M_{S})^2})^{1/3}\frac{M_{P}\sin i}{\sqrt{1-e^2}}. 
\end{equation}
In these equations $\theta$ is the true anomaly, $\omega_{*}$ is the argument of periastron, $e$ is eccentricity of the orbit, $P$ the orbital period of the planet in seconds, $M_{P}$ the mass of the planet in kilograms, and $M_{S}$ the stellar mass in kilograms.

The radial velocity at a given orbital phase is calculated and added to the appropriate spectrum. For this process the user has to provide the stellar mass, planetary mass, eccentricity and period of the planetary orbit (or semi-major axis). As an example, Fig.~\ref{Jitter_example}c shows the calculated radial velocity curve for a Neptune mass planet (17 Earth masses) around a solar mass star on a circular orbit with orbital period of 25 days. The radial velocities caused by the planet are calculated at the input rotational phases and shifts are introduced to the spectra. After this the spectra are re-analysed using the same cross-correlation method as for the case only containing the spot jitter. The resulting radial velocity curve of the spot and planet together is shown in  Fig.~\ref{Jitter_example}d. For this example, the full amplitude of the radial velocity variation is the same as from only the spot, 14.95 m/s, but the shape of the curve is very different.
 
\subsection{Testing the methods}

For testing the behaviour of the code a 5 degree radius equatorial solar-like spot was used (see Fig.~\ref{Jitter_example}a). The radius of the whole spot (umbra+penumbra) is 5$^{\circ}$, the radius of the umbra alone is 3$^{\circ}$. In some tests also a larger spot with the full radius of 17$^{\circ}$ and umbral radius of 10$^{\circ}$, is used. Both spots, which are used separately, are located at the equator and the temperature of the umbra is 4000~K, the penumbra is 4900~K and the unspotted temperature is 5800~K. All the tests were carried out using the grid size $60\times120$ (except the grid size tests). The spectra were created using spectral resolution of 100,000 (except in the resolution tests) and using 2.5 pixel sampling over one resolution element. The length of the wavelength strip used in one jitter calculation was always 46~{\AA}. The local line-profiles were calculated with $\pm 2$~{\AA} from the ends of the wavelength strip, to allow for large $v\sin i$ values in the calculations (i.e., the length of the spectrum for which local line-profiles were calculated was always 50~{\AA}).

\subsubsection{Recovering the planetary signal}

The accuracy at which the planetary signal can be recovered was tested using a smooth, i.e., unspotted, stellar surface. Solar-like configuration with effective temperature of 5800~K and $v\sin i$=~2.1 were used in the calculations. An Earth mass planet on a 25 day orbit around the one solar mass star was introduced to the spectra.

\begin{figure}
  \includegraphics[width=8cm]{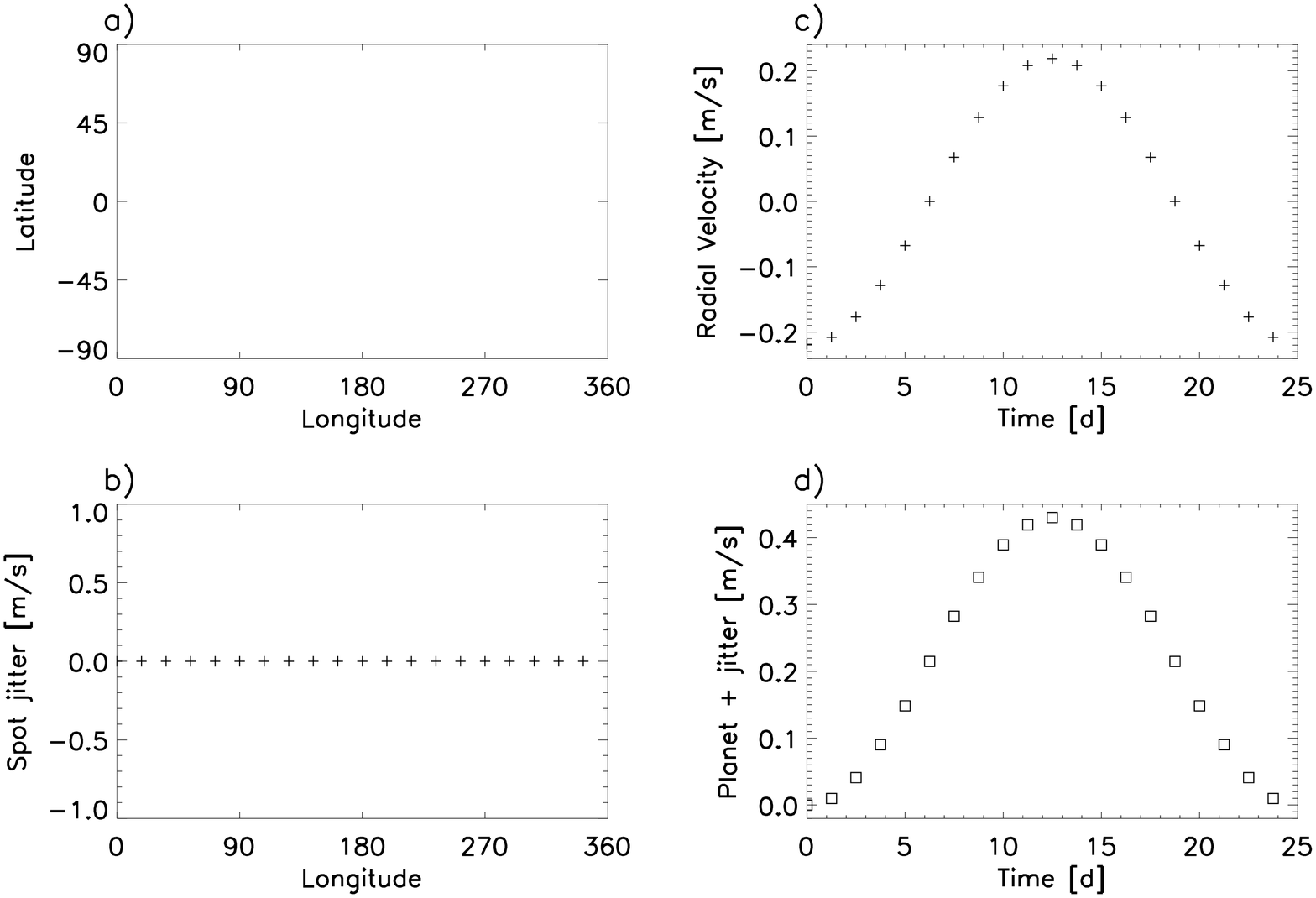}
  \caption{An example of recovering the planetary signal from the spectra. a) Stellar surface configuration without spots and surface temperature of 5800~K. This configuration was used for calculating the spectra for the planet recovery test. The x-axis gives the longitude in degree and y-axis the latitude, also in degrees. b) The resulting jitter curve from the unspotted surface. The x-axis is the longitude in degrees and y-axis the measured radial velocity jitter in m/s. The jitter values are calculated at 20 evenly spaced rotational phases on the wavelength region 5952--5998~{\AA}. c) Radial velocity curve of an Earth mass planet on a 25 day circular orbit around one solar mass star. The x-axis gives the time in days and y-axis the radial velocity in m/s. d) The radial velocity curve from the unspotted surface and an Earth sized planet. The x-axis gives the time in days and y-axis the measured radial velocity in m/s.
}
  \label{Planet_test}
\end{figure}

The results of the test can be seen in Fig.~\ref{Planet_test}. The jitter curve calculated from the unspotted stellar surface (Fig.~\ref{Planet_test}b) shows zero values for all the jitter measurements, as expected. The input curve of the Earth mass planet (Fig.~\ref{Planet_test}c) on the other hand is very well recovered from the spectra (Fig.~\ref{Planet_test}d). As discussed in Section~\ref{section_cc}, the radial velocity curve calculated from the spectra is normalised to the first measurement, explaining the different absolute values for the input and calculated radial velocities. The full amplitude of the jitter is almost identical in the full amplitude of the input radial velocity curve (0.4373~m/s) and the calculated one (0.4299~m/s). Out tests using other planetary masses show that the calculated full amplitude tends to be underestimated by $\sim$1.7\% in comparison to the real input radial velocity curves full amplitude.

\subsubsection{Effect of spectral resolution}

Radial velocity measurements are typically done using high resolution spectrographs with resolving power ($\lambda /\Delta\lambda$) 50,000--110,000. To test the effect of the spectral resolution on the jitter we calculated spectra using different resolving powers between 10,000 and 300,000 and the same spot configuration as shown in Fig.~\ref{Jitter_example}a. One resolution element always spans 2.5 wavelength steps, i.e., 'pixels'.

\begin{figure}
  \includegraphics[width=8cm]{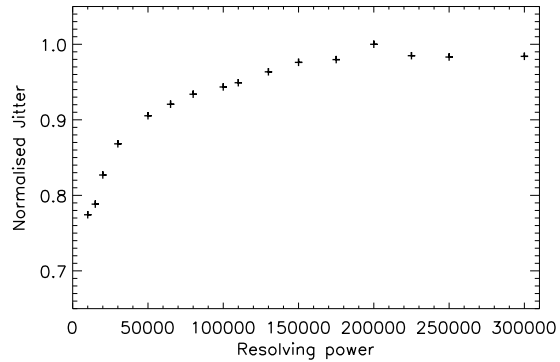}
  \caption{The spot-caused jitter with different spectrograph resolving power ($\lambda /\Delta\lambda$). The results are normalised to the maximum jitter case (i.e., resolving power of 150,000). The spot configuration used is the same as in Fig.~\ref{Jitter_example}a. In the plot the x-axis gives the resolving power and y-axis the jitter in m/s. The jitter increases with increasing resolving power.
}
  \label{Jitter_res}
\end{figure}

As can be seen from Fig.~\ref{Jitter_res} the full amplitude of the jitter decreases with decreasing spectrograph resolution, as has also been reported by other authors (e.g., \citealt{Desort07}). The results are normalised to the highest jitter, i.e., resolution 200,000 result of 15.8~m/s. Two different regimes can be seen in the jitter behaviour. Throughout the resolution range normally used for exoplanet searches, i.e., 50,000--300,000, the jitter remains at the level of about 90\% of the highest jitter value. The smallest jitter, about 77\% of the highest values, is seen at lowest spectral resolutions used in this test. 

This behaviour can be explained by the spot contribution on the line-profile shape being better resolved at high spectral resolution and getting more and more diluted with decreasing spectral resolution. Still, the effect of the spectrograph resolution in the jitter amplitude is not very strong and the accuracy of the radial velocity measurements decreases with the decreasing spectral resolution. 

\subsubsection{Effect of the width of the spectral region}

In real high precision spectroscopic observations the whole optical wavelength range is typically used for determining the radial velocity. Still, as there is no noise added to the spectra the increase in the width of the spectral range should not have a major influence on the results, if wide enough wavelength region is used. To test this assumption we have calculated jitter from seven different wavelength ranges, with the width spanning from 10 {\AA} to 70~{\AA}. The test uses wavelengths between 5925 {\AA} and 6000~{\AA}, and spectral resolution of 100,000.

\begin{figure}
  \includegraphics[width=8cm]{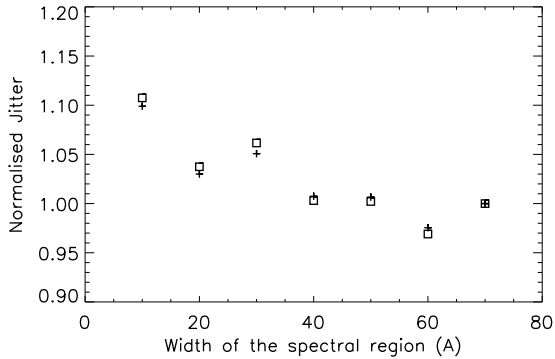}
  \caption{The spot-caused jitter with different widths of the spectral region. Two different spot configurations were used for this test: the same as in Fig.~\ref{Jitter_example}a shown by plus-signs, and one with the full spot radius of 17$^{\circ}$ denoted by squares. The results for both spot configurations are normalised individually to the results from the widest spectral region (70~{\AA} case). In the plot the x-axis gives the width of the spectral region in {\AA}nsgtr{\"o}m and y-axis the jitter in m/s.
}
  \label{Jitter_range}
\end{figure}

Figure~\ref{Jitter_range} shows the normalised jitter for two different spot configurations: 5 degree spot (plus-signs) and 17 degree spot (squares). As expected, the jitter is highest for the 10~{\AA} wide wavelength range, and decreases slightly when going to the 40~{\AA} wide wavelength region. The wider wavelength regions, from 40~{\AA} to 70~{\AA}, all show similarly small jitter, and the results from more narrow spectral ranges have 5--10\% larger jitter values. Still, we cannot say based on this test, whether the jitter would decrease further if significantly wider wavelength ranges would be used, but for limiting the calculation times, 46~{\AA} wide wavelength region is used throughout this paper.

\subsubsection{Size of the spatial grid}

Testing which impact the grid size, i.e., spatial resolution on the stellar surface, has on the jitter is difficult. The location and fractional size of the spots have to be kept identical throughout the test, which is of course strictly speaking impossible to do. 

For the test a random spot configuration with a grid size $20 \times 40$ was created. The input map has a spot filling-factor of 1.5\% and the spots were occurring at latitude range -30$^{\circ}$ -- +30$^{\circ}$. The grid resolution was increased and filling-factor kept as close to the original as possible with the increasing number of grid elements. The scalings did not have a significant effect on the jitter and the filling factor. The jitter values were between 24.6 and 29.0 m/s with all the grid sizes. These tests imply that the results are not critically dependent on the grid size, but to allow for also small spots on the surface we have used the grid size $60\times 120$ throughout this paper.

\subsubsection{Signal-to-noise ratio of the observations}

To investigate the effect of noise on the radial velocity curves we add Gaussian noise of a specified signal-to-noise ratio (S/N) to the spectra. The level of the continuum represents signal-level while the standard deviation of the noise represents the noise-level. Since the continuum is normalised $S/N=\frac{signal~level}{noise~level}=\frac{1}{\sigma}$, leading to $\sigma=\frac{1}{S/N}$. A sequence of pseudo-random numbers are generated using IDL routine RANDOMN, which creates Gaussian random numbers using Box-Muller method. The noise is then applied to the spectra.

\begin{figure}
  \includegraphics[width=8cm]{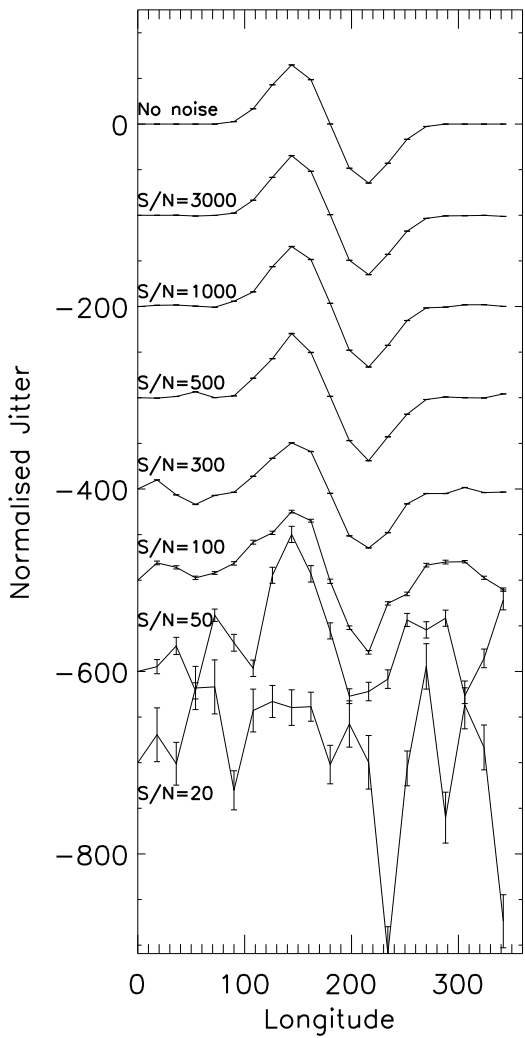}
  \caption{Jitter curves caused by a 17$^{\circ}$ radius spot with different signal-to-noise ratios of the spectra. In the plot the x-axis gives the stellar longitude in degrees and y-axis the jitter in m/s. The topmost jitter curve is from the case without noise, and the other jitter curves are off-set from this curve to show the different behaviour better. The S/N values are written on the plot for each jitter curve. The plot also shows the error of each jitter value. The error has been calculated as described in Section~\ref{section_cc}.
}
  \label{Jitter_SN}
\end{figure}

For testing the effect of noise in the jitter curves we use spectra calculated for a single equatorial spot with a full radius of 17$^{\circ}$. Jitter curves are calculated for seven different cases, one with no noise and six with different S/N values ranging from 20 to 3000. The resulting curves are shown in Fig.~\ref{Jitter_SN}. The symbols used in the plot also show the errors of the individual jitter values (described in Section~\ref{section_cc}). The shape of the jitter curve is easily recognisable until S/N less than 100. The general shape can still be recovered from the spectra with S/N=50, but with S/N=20 the shape becomes basically unrecognisable.

We want to still note that each spectral line represents an individual measurement of the Doppler shift of the star. If a total of $N$ lines are used for the Doppler measurement, then the error will be decreased by a factor of $\sqrt{N}$ over a single line measurement. In the tests carried out here, short wavelength ranges of 50~{\AA} are used. In the solar-like case these wavelength regions have ~30--100 spectral lines, the exact number depending on the wavelength (typically more lines in the blue part and less in the red). Planet searches on the other hand use echelle spectra with a few thousand lines in them. Therefore, if our tests only have at most a 10 fold gain over single line measurements, the real planet searches usually have approximately 40 fold gain. For this reason in the following investigations we will use the calculated spectra without added noise. This will enable us also to study the ideal detection cases.

\section{Results}

We apply the developed codes to study the starspot jitter in solar-like stars. First, general properties of jitter are studied and compared to the earlier published results by other groups. Afterwards, two different activity cases are investigated: solar-like low activity and very active young solar analogues. In these investigations solar-like temperatures (unspotted surface 5800~K, umbra 4000~K and penumbra 4900~K) are used. If not mentioned otherwise, the gird size is $60\times 120$, the wavelength region is 5952--5998~{\AA} and the spectrograph resolving power ($\lambda /\Delta\lambda$) is 100,000 with 2.5 pixel sampling. Similarly, the inclination was set to 90$^{\circ}$ and $v\sin i$ was fixed to 2.1~km/s (if not mentioned otherwise). 

\subsection{General properties}

\subsubsection{Jitter with wavelength}

Several works have already shown that the spot caused jitter decreases with increasing wavelength (e.g., \citealt{Desort07}; \citealt{Reiners10}). In a recent paper \cite{Marchwinski} show that, based on solar observations, near-infrared has lower estimated radial velocity jitter throughout the entire solar cycle than the optical wavelengths have. Here tests using spectra with lengths of 46~{\AA} at different wavelength regions between 3700~{\AA} and 9100~{\AA} are carried out. The spectrograph resolving power ($\lambda /\Delta\lambda$) is kept constant at 100,000, by changing the $\Delta\lambda$ according to the wavelength region. Two spot sizes, 5$^\circ$ and 17$^\circ$ full spot radius (umbra+penumbra), are used. The radii of the umbra are 3$^\circ$ and 10$^\circ$, respectively. 

\begin{figure}
  \includegraphics[width=8cm]{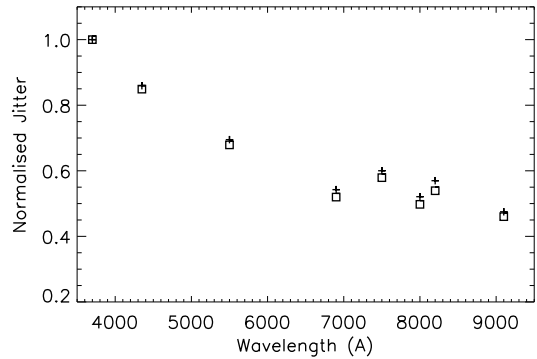}
  \caption{Full amplitude of the spot-caused jitter at different wavelengths. Two different spot configurations were used for this test: the same as in Fig.~\ref{Jitter_example}a shown by plus-signs, and one with the full spot radius of 17$^{\circ}$ denoted by squares. The results are normalised to the highest jitter case of 3700~{\AA}. In the plot the x-axis gives the wavelength in {\AA}ngstr{\"o}m and y-axis the normalised jitter. As has been seen in other studies too, the jitter decreases towards the longer wavelengths, but also the scatter in the jitter values increases.
}
  \label{Jitter_wl}
\end{figure}

The results of the jitter investigation over the wavelength are shown in Fig.~\ref{Jitter_wl}. The results are normalised to the highest jitter case, that of 3700~{\AA}. The results for the 5$^{\circ}$ spot are shown by plus-signs and for the 17$^{\circ}$ spots by squares. As can be seen, the normalised values and the over-all behaviour are very similar for both spot sizes. The absolute values are naturally very different: the full amplitude of the jitter for the 5$^{\circ}$ spot at 3702--3748~{\AA} is 23.7~m/s and for the 17$^{\circ}$ spot 213.7~m/s. 

The jitter decreases by about 50\% between the wavelengths 3700{\AA} and  7000~{\AA}. At longer wavelengths than this the reduction in jitter amplitude is not as clear as before, and the whole behaviour becomes more chaotic. The increased scatter at red wavelengths could possibly be due to varying number of spectral lines in the spectral windows used in this analysis. In general, there are less spectral lines at the red wavelengths and therefore different red regions can have very different number of spectral lines. Investigation extending to longer wavelengths would be needed to study whether or not the decrease continues to infrared. A similar plateau in the wavelength dependence of the jitter around 8000--10,000~{\AA} is seen for solar-like stars by \cite{Reiners10}. In their work some further decrease in the jitter amplitude is seen in the infrared wavelengths, as is also detected by \cite{Marchwinski}.

\subsubsection{Jitter with $v\sin i$}

It has been shown before that the jitter depends strongly on the stellar rotation rate (e.g., \citealt{SaarDonahue97}; \citealt{Desort07}; \citealt{Boisse12}). Also in this work the effect was studied and compared to the previously published results. Two spot cases, with the full spot radius of 5$^{\circ}$ and 17$^{\circ}$, were used in the investigation.

\begin{figure}
  \includegraphics[width=8cm]{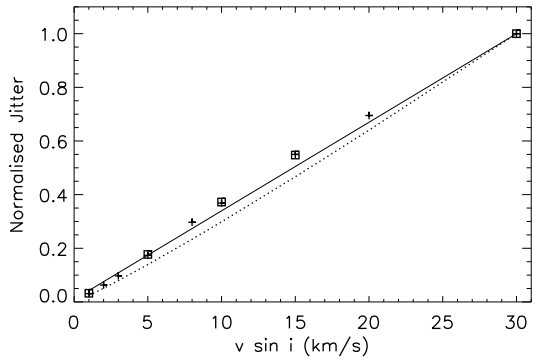}
  \caption{Full amplitude of the spot-caused jitter with different stellar rotation rates. Two different spot configurations were used for this test: 5$^{\circ}$ radius equatorial spot shown by plus-signs, and 17$^{\circ}$ radius spot denoted by squares. The results are normalised to the highest jitter case of $v\sin i = 30$~km/s. In the plot the x-axis gives the $v\sin i$ in km/s and y-axis the normalised jitter. The solid curve is the fit to the jitter obtained from the 5$^{\circ}$ spot, and the dotted line is the $v\sin i$ dependence law deduced by \citet{Desort07}.
}
  \label{Jitter_vsini}
\end{figure}

Figure~\ref{Jitter_vsini} shows the results from the two spot cases for $v\sin i$ values ranging between 1~km/s and 30~km/s. The results are normalised to the largest jitter case of 30~km/s. The results from the calculations using 5$^{\circ}$ spot are shown with plus-signs, and the ones from 17$^{\circ}$ spot with squares. It is clear that the jitter increases with increasing $v\sin i$, as has been shown by earlier studies (e.g., \citealt{SaarDonahue97}; \citealt{Desort07}). In addition, when the results are normalised to the highest value the values calculated from the two different spot sizes show the same trends. This means that the magnitude of the jitter depends on the spot size, but the increase in $v\sin i$ affects the jitter measurement the same way regardless of the spot size. 

\citet{Desort07} published a formula for jitter--$v\sin i$ correlation. They obtained from their tests that the jitter amplitude depends on the spot coverage of the visible surface and $v\sin i$. We use their formula to compute jitter--$v\sin i$ dependence, and compare it to our results. The resulting linear trend is plotted in Fig.~\ref{Jitter_vsini} with a dotted line. Our results agree well with those obtained by \citet{Desort07}. The amplitude of the jitter estimated by \citet{Desort07} is very similar to the ones given by \citet{SaarDonahue97} and \citet{Boisse12}. On the other hand, when comparing our jitter amplitudes to the ones from \citet{Desort07}, our values are larger. The jitter from the five degree spot is approximately 20\% lower at the low $v\sin i$ values in the results obtained using the formula by \citet{Desort07}. The situation improves towards the higher $v\sin i$ values, and for the $v\sin i = 30$km/s the difference in only few per cent. Some of the discrepancy could be explained by different wavelength regions and spectral resolutions that were used in these investigations.

\subsubsection{Effect of inclination}

\begin{figure}
  \includegraphics[width=8cm]{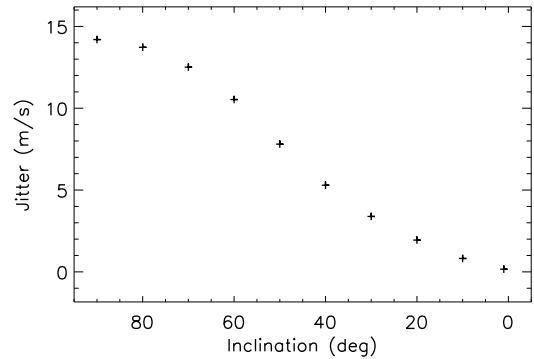}
  \caption{Full amplitude of the spot-caused jitter at different inclinations of the stellar rotation axis. The spot configuration is the same as in Fig.~\ref{Jitter_example}a.  In the plot the x-axis gives the inclination in degrees and y-axis the full jitter amplitude in m/s. As is expected the level of jitter lowers when the visibility of the equatorial spot decreases with decreasing inclination angle (viewing progressively more pole-on).
}
  \label{Jitter_inc}
\end{figure}

The effect of inclination of the stellar rotation axis was studied using a fixed $v\sin i$ of 2~km/s. When the inclination is changed, the $v\sin i$ changes too. If a star is viewed pole-on there would be no rotational broadening. As the jitter also depends on the broadening of the spectral lines, we decided to use a fixed $v\sin i$ value for this test. The full amplitude of the measured jitter at different inclination angles is shown in Fig.~\ref{Jitter_inc}. The test reveals the expected behaviour of the jitter, where the amplitude decreases with the decreasing visibility of the equatorial spot (decreasing inclination angle). The equatorial spot has the maximum effect on the jitter when the star is viewed equator on. The visibility of the spot is reduced when viewed increasingly from the direction of the pole, and thus also the impact of the spot on the jitter amplitude decreases. When viewed from almost the pole (inclination of 1 degree), the spot is seen at the limb and at all the rotational phases. This test does not include the effect of decreasing line broadening with decreasing inclination, which would make the change in jitter amplitude even more pronounced. 

\subsection{Solar-like activity patterns}

For investigating the typical jitter amplitude caused by solar activity we have created 50 random spot configurations. All the spot configurations have spot filling factors of 0.1\%. This value is over the whole stellar surface and represents normal solar activity level (see, e.g., \citealt{Balmaceda09}). The spots have been restricted to occur between latitudes -30$^{\circ}$ and +30$^{\circ}$, which also is the typical latitude range for solar activity. 

\begin{figure}
  \includegraphics[width=8cm]{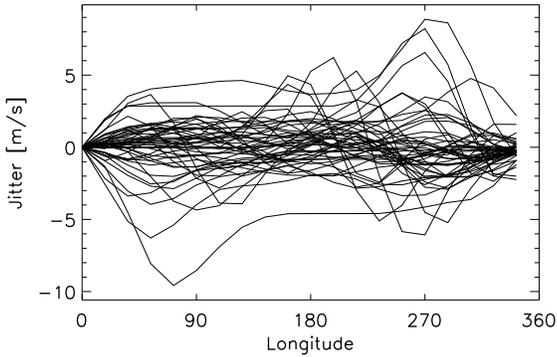}
  \caption{Jitter curves calculated from 50 different randomly created solar-like spot configurations with the spot filling factors of 0.1\% and spot latitudes restricted to -30$^{\circ}$ and +30$^{\circ}$. The x-axis is the longitude in degrees and y-axis the jitter in m/s. Note that the curves are created in such a way that the jitter at first observations (first longitude) is always zero.}
  \label{Solar_jitter}
\end{figure}

The jitter curves calculated from all the 50 different spot configurations are shown in Fig.~\ref{Solar_jitter}. As can be seen the typical jitter with this spot configuration varies between -2m/s and +2m/s. The mean amplitude of the full jitter is 4.5~m/s with the standard deviation being 2.8~m/s. The minimum full amplitude from these spot configurations is 1.5~m/s and the maximum 12.3~m/s. 

\begin{figure}
  \includegraphics[width=7cm]{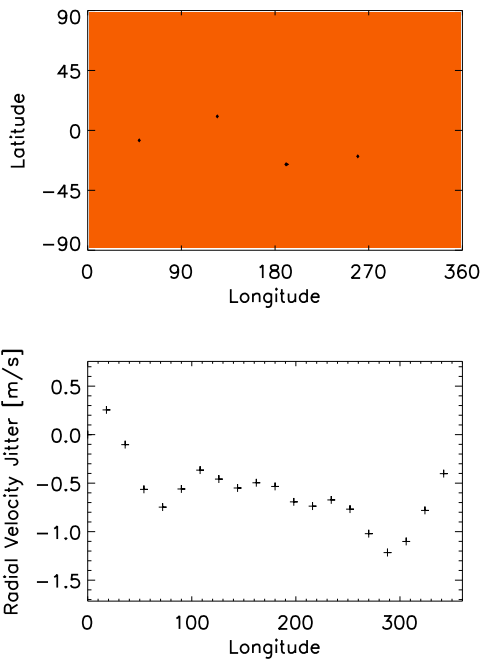}
  \caption{Spot configuration and jitter curve from the spot configuration that out of the 50 randomly created configurations results in the smallest jitter. The jitter has a full amplitude of 1.5~m/s. The upper plot gives the spot configuration. The x-axis is the longitude and y-axis the latitude, both are given in degrees. The lower plot shows the corresponding jitter curve. Here the x-axis is the longitude in degrees and y-axis jitter in m/s.
}
  \label{Solar_jitter_min}
\end{figure}

\begin{figure}
  \includegraphics[width=7cm]{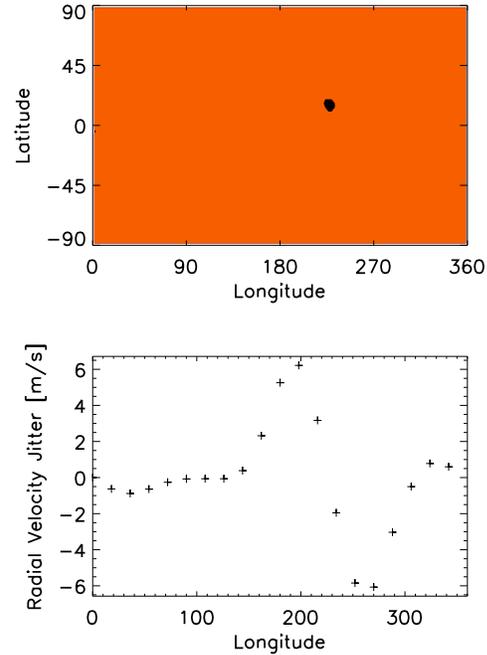}
  \caption{The same as fig.~\ref{Solar_jitter_min}, but now for the largest jitter case. The resulting jitter has full amplitude of 12.3~m/s. 
}
  \label{Solar_jitter_max}
\end{figure}

The spot configurations and resulting jitter are plotted in Fig.~\ref{Solar_jitter_min} and Fig.~\ref{Solar_jitter_max} for the minimum and maximum jitter case, respectively. The smallest jitter results from a case where several tiny spots are distributed relatively evenly over the longitude. On the other hand, the largest jitter arises from a spot configuration where there is one large spot together with one tiny one (at longitude 0, just below the equator). This is what one would expect and means that the exact jitter from a solar-like spot configuration depends largely on the exact spot distribution and how concentrated into active regions the spots are. This is in line with the studies which show that the radial velocity jitter can be estimated based on the photometric variability (e.g., \citealt{Aigrain12}). Concentrated spots introduce more photometric variability, and also more radial velocity jitter.

We have to note though, that due to the limited resolution of the grid on the stellar surface, the filling factor is not always exactly the same. The code for creating the spotted surface will add a spot and then check the filling factor. Another spot is added if needed. This is done until the filling factor is greater than, or equal to, the one that was specified. Therefore the generated filling factors can be slightly above the input value, and are not always exactly identical. 

\subsection{Active solar-like stars}

For investigating what the jitter behaviour of young, very active solar-like stars would be, we use the temperature maps of V889~Her by \citet{Jarvinen08}. Their observations of the photospheric and chromospheric properties of V899~Her indicate that the quiet photosphere of V889~Her is similar to the one of the present day Sun, while the chromosphere shows much stronger activity. Their temperature maps, obtained using Doppler imaging, show that the polar regions are covered by spots, which are about 1500 K cooler than the quiet photosphere. Some evidence for cyclic magnetic activity is also seen both from photometry and Doppler imaging results.

Here we calculate the jitter resulting from the temperature maps of V889~Her for four different years: 1999, 2001, 2005, and 2007 (\citealt{Jarvinen08}). In the jitter calculations the wavelength range 5952--5998~{\AA} is used and the inclination and $v\sin i$ are set to the ones determined from Doppler imaging, $i=60^{\circ}$ and 37.5~km/s, respectively. The spectral resolution used in the calculation was the same as the one in the original observations (77,000 and 2.5 pixels over the resolution element) and the grid size of the visible stellar surface is set to that of the original temperature map, $30 \times 60$.

\begin{figure*}
  \includegraphics[width=16cm]{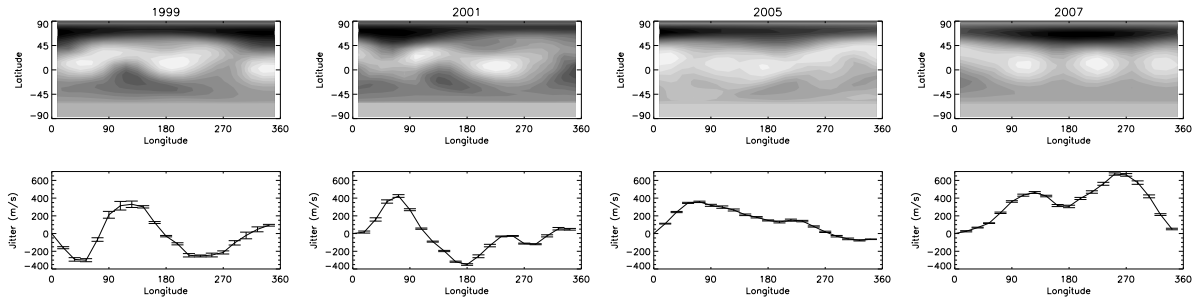}
  \caption{Temperature maps of V889~Her \citep{Jarvinen08} and the resulting jitter. The maps shown in the upper part of the plot are from years 1999, 2001, 2005, and 2007 (from left to right). The temperature ranges in the maps are: 4624--6327~K (1999), 5251--6037~K (2001), 4498--6185~K (2005), and 4561--6268~K (2007). The x-axis gives the longitude in degrees and y-axis the latitude in degrees. The lower panels in the plots show the calculated jitter from each temperature map at 20 different phases evenly distributed over the stellar rotational phase. The x-axes give the longitude in degrees and y-axes the jitter in m/s.
}
  \label{Jitter_V889Her}
\end{figure*}

The original V889~Her temperature maps and the calculated jitter are shown in Fig.~\ref{Jitter_V889Her}. The full amplitude of jitter varies between 435 m/s calculated from the 2005 map and 774 m/s obtained from the 2001 map. These values are similar to radial velocity variations caused by a hot Jupiter around solar mass star. \citet{Moulds13} show that in this kind of cases some of the activity signal can be cleaned from the spectral line profiles and Jupiter mass planets on close orbits can be recovered. In a recent paper \cite{Jeffers14} study the detectability of planets around young active solar-like stars. They conclude that Jupiter-mass planets can be detected on close-in orbits around fast-rotating young active stars, Neptune-mass planets around moderate rotators and Super-Earths only around very slowly rotating stars. The calculations carried out here based on the V889~Her spot configurations support these conclusions.

\section{Discussion}

\begin{figure*}
  \includegraphics[width=16cm]{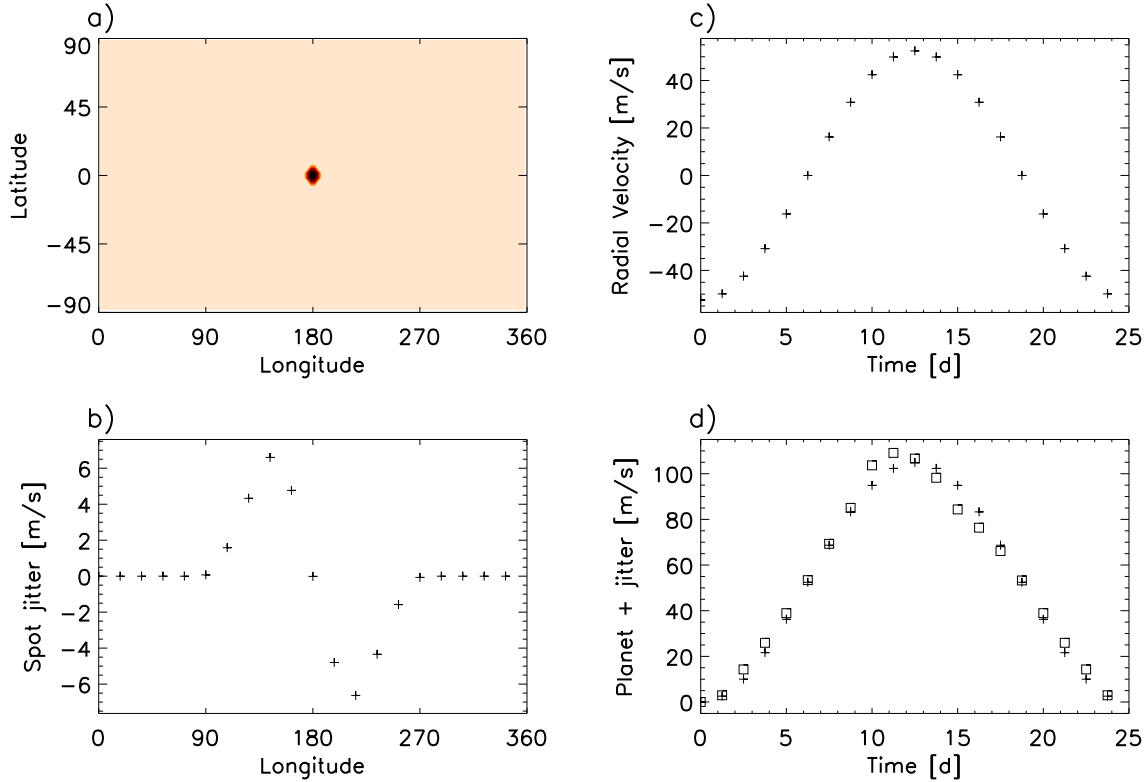}
  \caption{An example of the effect of the spot signal on the radial velocity curve of the planet. The axes in the subplots are the same as in Fig.~\ref{Planet_test} a) Stellar surface configuration with one equatorial 5 degree spot of 4000~K and unspotted surface temperature of 5800~K. b) The resulting jitter curve from the spot configuration. The jitter values are calculated at 20 evenly spaced rotational phases from the wavelength region 5952--5998~{\AA} and spectral resolution of 100,000. c) Radial velocity curve of an 0.8 Jupiter mass planet on a 25 day circular orbit around one solar mass star. d) The radial velocity curve from the spotted surface and the planet is given with squares. The plus-signs give the original 'planet only' radial velocity curve (also shown in subplot c).}
  \label{Planet_e}
\end{figure*}

\subsection{Starspots as a confusion factor}

Not only can starspots cause noise in the radial velocity measurements, they can also mimic planetary signals and change the shape of the radial velocity curve. 

For example, a large spot group close to the pole, which is viewed at low inclination, is visible on the surface all the time. The spot group causes larger jitter when it has a higher visibility, i.e., in the front, than when it is behind the pole close to the limb of the star. This behaviour can be confused with the radial velocity variation caused by an orbiting planet and Keplerian fits to the curve can be easily obtained. Several different cases of this kind of confusion have been already discussed by \citet{Desort07}.

Another confusion occurs when the location of the spot on the surface is such that together with the planetary signal it actually causes subtle changes in the shape of the measured radial velocity curve. For example spots at `correct' location on the surface can change the shape of the radial velocity curve of a circular orbit into something that could be interpreted as a more eccentric orbit. An example of this kind of changes is given in Fig.~\ref{Planet_e}. There the 5 degree spot from our tests (see Fig.~\ref{Jitter_example}a) is used together with a 0.8 Jupiter mass planet on a circular orbit. The calculations are carried out at wavelength 5952--5998~{\AA} and using spectral resolution of 100,000. The spot itself introduces jitter (see Fig.~\ref{Planet_e}b) which is only about 10\% of the radial velocity variation of the planet (Fig.~\ref{Planet_e}c). Still, the combination of these two changes the input radial velocity curve in subtle ways. In Fig.~\ref{Planet_e}d the total radial velocity curve from the planet and spot jitter is plotted with squares. If that is compared to the original planet radial velocity curve, which is overplotted with plus-signs, one can see that the phase of the maximum radial velocity has changed. The slope of the increase is now steeper and the declining slope more shallow. These changes would be interpreted as slightly eccentric orbit instead of a circular one. Naturally the activity could also work the other way, and make eccentric orbits appear more circular.

\subsection{Solar-like activity cycle}

One can question how typical the solar activity level is among solar-like stars in general. There have been early tentative suggestions that the Sun might be photometrically more quiet than similar stars (e.g., \citealt{Radick98}). The high precision photometric data from Kepler and CoRoT can help to answer this question. The early results from Kepler implied that the Sun could indeed be more quiet than an average solar-type star (\citealt{McQuillan12}; \citealt{Gilliland11}). A new study by \cite{Basri13} has revisited the activity fraction of solar-like stars in the Kepler data. Their results show that 25--30\% of solar-type stars are more active than the Sun. The exact fraction depends on the timescales used in the study, what is meant by `more active than the Sun', and on the magnitude limit of the sample. In light of these investigations it seems appropriate to use the solar cycle as a proxy for cyclic activity in other stars.

\subsubsection{Jitter during activity cycle}

The number of spots on the solar surface varies with the 11 year cycle. The spot coverage of the solar surface varies from zero to about 0.5\%. Typically the minima show very small spot coverage factors of 0.01--0.02\%, or even zero for extended periods. The maxima on the other hand have quite varying spot coverage fractions. A small maximum has a typical spot coverage of 0.2\%, whereas a strong maximum has a spot coverage fraction of 0.5\% (see, e.g., \citealt{Balmaceda09}). During the cycle the latitudes of the spots also change. The new cycle starts with a small amount of spots which appear at relatively high latitudes, around $\pm30^{\circ}$. During the cycle the activity migrates slowly towards the equator, and during the next minimum the last spots of the old cycle appear at latitudes $\pm10^{\circ}$. This behaviour also affects the activity-caused jitter. For more details on the latitudinal migration of sunspots within the solar cycle see, e.g., \cite{Carrington1858}, \cite{Maunder1903}, and \cite{Hathaway11}.

We have created random spot configurations with typical sunspot coverage fractions and latitude ranges for studying the jitter over the solar-like activity cycle. In total six different activity cases were investigated: two activity minima, one average activity case and three activity maxima. The details of the cases that were studied are given in Table.~\ref{cycle_jitter_table}. For all the cases 100 different spot configurations were created. The jitter was calculated at wavelengths 5952--5998~{\AA} using a resolution of 100,000, inclination of 90$^{\circ}$, and $v\sin i$ of 2.1~km/s..

\begin{table}
  \centering
    \caption{The different activity cases created to study the behaviour of the spot caused jitter over solar-like activity cycle.}
  \begin{tabular}{@{}lcccc@{}}
\hline
Case  & spot frac. & latitude & jitter & $\sigma$ \\
      &               &          & [m/s]       & [m/s] \\
\hline
Minimum 1 & 0.02\% & +20 -- +30$^{\circ}$ & 1.8 & 1.6 \\
Minimum 2 & 0.02\% & -10 -- +10$^{\circ}$ & 2.0 & 1.3 \\
Average & 0.1\% & -20 -- +20$^{\circ}$    & 4.2  & 2.4  \\
Maximum, small & 0.2\% & -20 -- +20$^{\circ}$  & 5.3 & 2.9 \\
Maximum, medium & 0.3\% & -20 -- +20$^{\circ}$ & 6.6 & 2.4 \\
Maximum, large & 0.5\% & -20 -- +20$^{\circ}$ & 7.7 & 2.9 \\
\hline
\end{tabular}
\label{cycle_jitter_table}
\end{table}

\begin{figure}
  \includegraphics[width=8cm]{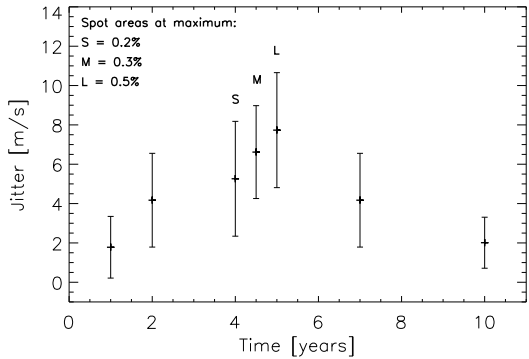}
  \caption{Results for the investigation of jitter during a solar-like activity cycle. The x-axis gives the time in years and y-axis the jitter in m/s. The mean full jitter amplitude calculated from 10 individual spot configurations with the different activity cases given in Table~\ref{cycle_jitter_table} are plotted approximately at the time they would occur during the solar 11 year cycle. The errorbar gives the standard deviation of the measurements from the 10 individual spot configurations.
}
  \label{Solar_jitter_cycle}
\end{figure}

The results from the jitter calculations are shown in Fig.~\ref{Solar_jitter_cycle}, and the mean jitter and its standard deviation are also given in Table~\ref{cycle_jitter_table}. In the plot, the x-axis gives the time in years, and the mean jitter of the different activity cases have been plotted at a time it would typically be observed during the solar 11 year cycle. The standard deviation of the jitter from 100 individual spot configurations has been plotted as an errorbar for each activity case. The result from the average jitter case has been plotted both in the rising and declining phase of the activity. 

For the maximum solar activity three different jitter cases were calculated. They are marked in Fig.~\ref{Solar_jitter_cycle} by letter S (small maximum, spot fraction 0.2\%), M (medium maximum, spot fraction 0.3\%) and L (large maximum, spot fraction 0.5\%). As expected, the large sized maximum results in the largest jitter values, and the small sized maximum induces the smallest jitter. Still, all the average jitter values for different solar maxima cases are the same within the standard deviation, and the variation in the jitter, as is seen from the standard deviation, is similar for all the cases.

Both of the minima cases have the smallest jitter values, and also the smallest standard deviation. This is because with such a small spot coverage (0.02\%) basically only one small spot is present on the surface. The exact location of the spot changes slightly, and the differences in latitude result in small changes in the full jitter amplitude. It is interesting to note that the mean jitter from the early cycle case, where the spots are around latitudes $\pm30^{\circ}$ results in smaller jitter than the late cycle case with spots at latitudes $\pm10^{\circ}$. The effect of the jitter is largest when the spots are best visible, i.e., in the case of inclination 90$^{\circ}$ around equator. One should also note that the spots for the early cycle case have been created with spot latitude 20$^{\circ}$--30$^{\circ}$, not taking into account that on the Sun the spots would appear both around latitude -30$^{\circ}$ and +30$^{\circ}$. Regardless, with such small spot coverage fractions only one spot is created, and therefore this has no practical effect in the full jitter amplitude, which is what is studied here.

These calculations do not take into account the lifetime of the sunspots. On the other hand, the larger sunspot on average live longer, couple of weeks, instead of couple of days (see, e.g., \citealt{Gnevyshev38}; \citealt{Waldmeier55}; \citealt{Petrovay97}). Here we by necessity concentrate on larger spots. Therefore, not taking into account the lifetime of the spots should not significantly affect our investigation. In addition, several authors have reported that the solar activity tends to occur at active longitudes (e.g., \citealt{Berdyugina03}; \citealt{Usoskin05}; \citealt{Juckett06}). This effect has not been taken into in our calculations; we have restricted the spot occurrence only in the latitude direction. The solar active longitudes would group the spots in longitudinal direction and increase the jitter. Thus, the random distribution investigated here can be considered as the lower limit for the jitter.

\subsection{Detecting planets on a one year orbit}

An interesting question is how the jitter during a solar-like cycle would affect the detection of a planet in the habitable zone of its host star. We have studied this issue by introducing a planet on one year orbit around a star showing a cycle similar to the one described in the previous Section. 

\begin{figure}
  \includegraphics[width=8cm]{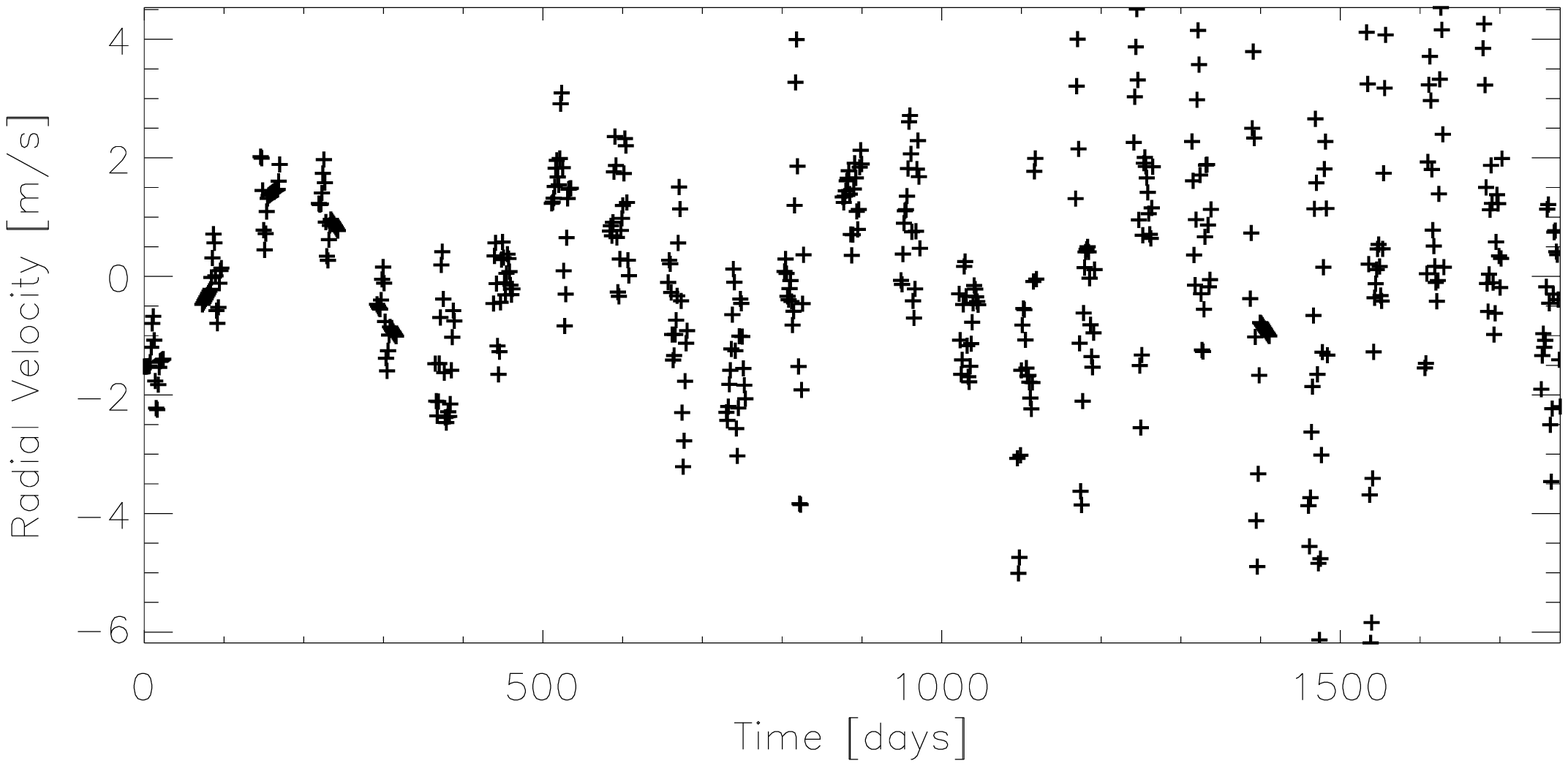}
  \caption{Simulated radial velocity measurements of Neptune mass planet orbiting a solar-like star with a solar-like activity cycle. The observations span five years, and during each year there are five separate observing runs lasting 25 nights. The x-axis gives time in days and y-axis the radial velocity in m/s.
}
  \label{Sun_Neptune}
\end{figure}

The cycle is thought to last 11 years, like in the Sun, and the observing period is five years, covering the cycle from minimum to maximum. Each year the target can be observed during its visibility, here called the observing season, and each season a fixed number of observing runs is carried out. The length of the observing run is always 25 days, which is also the rotation period of the target star. During one run 20 evenly spaced observations are carried out. For each observing run during the first year a spot configuration from the solar minimum case is randomly chosen. For the second and third year the spot configurations have been chosen randomly from the average activity case. During the fourth year spot configurations are from the small activity maximum case, and during the fifth year from the medium maximum maps. After this a planet on a one year orbit is introduced to the jitter measurements. Fig.~\ref{Sun_Neptune} shows an example of radial velocity measurements created this way. In the example the planet has the same mass as Neptune (17 Earth masses) and the observing season lasts the whole year and has five individual observing runs during it. It can easily be seen that the activity caused noise in the radial velocity measurements increases with the advancing activity cycle.

\begin{figure}
  \includegraphics[width=8cm]{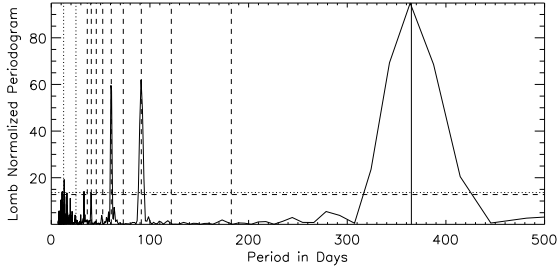}
  \caption{Lomb-Scargle periodogram obtained from the simulated radial velocity measurements presented in Fig.~\ref{Sun_Neptune}. The dotted horizontal line is the analytical 3$\sigma$ detection threshold and the dashed horizontal line the numerical one. The solid vertical line gives the original period of the planet (365 days) and the dashed vertical lines give its harmonics. The dotted vertical lines are the rotation period of the star (25 days) and half of the rotation period of the star. The x-axis gives the period in days and y-axis the power spectral density.
}
  \label{Sun_Neptune_Scargle}
\end{figure}

\begin{figure}
  \includegraphics[width=8cm]{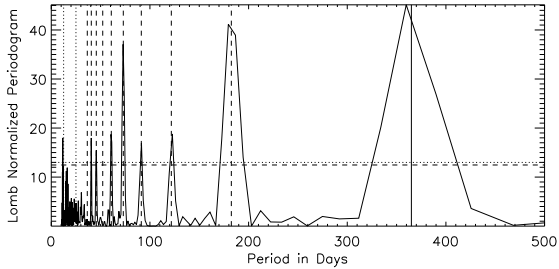}
  \caption{The same as Fig.~ \ref{Sun_Neptune_Scargle}, but now with observing season only lasting 150 days and with only three observing runs during this time.
}
  \label{Sun_Neptune_Scargle150}
\end{figure}

For investigating the detectability of the planet we use Lomb-Scargle period search method for unevenly sampled data (\citealt{Scargle82}; \citealt{Horne86}; \citealt{Press89}). The periodogram resulting from the simulated data presented in Fig.~\ref{Sun_Neptune} is shown in Fig.~\ref{Sun_Neptune_Scargle}. The dominant frequencies in this case are the orbital period of the planet (marked by a solid vertical line) and its harmonics (marked by dashed vertical lines). The dotted vertical lines denote the rotation period (25 days) and half a rotation period of the star, whereas the horizontal dotted line is the analytical 3$\sigma$ detection threshold following the false alarm probability (FAP) formulation of \citet{Scargle82} and taking into account the modifications of \citet{Horne86}. The horizontal dashed line, on the other hand, is the FAP obtained from white noise simulations with 10000 iterations. The two FAP values are similar, which is to be expected because our data is not severely unevenly distributed, the case where the analytical method would strongly underestimate the FAP. Our data are evenly distributed over the observing run, and the observing runs are evenly distributed throughout the year. In this simulation the Neptune mass planet is easily detected even with the noise from the stellar activity. The case remains the same if we shorten the observing season to 150 days and only have three observing runs during it. This is a more realistic case because of the limited visibility of the targets and also owing to the telescope time allocation process. The Lomb-Scargle periodogram of this case is shown in Fig.~\ref{Sun_Neptune_Scargle150}. The orbital period of the planet is still easily detectable, but the harmonics of the orbital period have become more pronounced. Part of the effect is also due to aliasing caused by larger data gaps.

\begin{figure}
  \includegraphics[width=8cm]{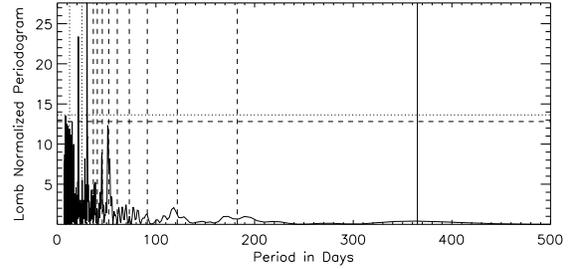}
  \caption{The same as Fig.~ \ref{Sun_Neptune_Scargle}, but now with an Earth-mass planet.
}
  \label{Sun_Earth_Scargle_runs5}
\end{figure}

\begin{figure}
  \includegraphics[width=8cm]{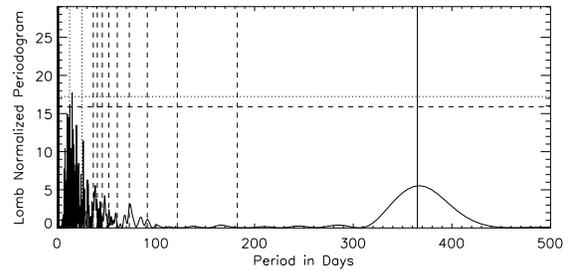}
  \caption{The same as Fig.~ \ref{Sun_Neptune_Scargle}, but now with an Earth-mass planet and 50 observing runs during the observing season of one year.
}
  \label{Sun_Earth_Scargle_runs50}
\end{figure}

On the other hand, the possibility of detecting Earth-mass planet around a solar-like star is much more challenging. The periodogram from the optimum case where the observing season lasts the whole year and there are five individual observing runs during the season is shown in Fig.~\ref{Sun_Earth_Scargle_runs5}. No indication of the true orbital period is seen, and the strongest periodicity is half the stellar rotation period. The situation is somewhat improved when the number of observing runs is increased. In Fig.~\ref{Sun_Earth_Scargle_runs50} a case where there are 50 observing runs during each year is shown (this implies observing more than once a night). In this case the signature of the true orbital period starts to emerge, but still it cannot be considered significant. Our tests show that the 3$\sigma$ detection limit for a planet in a habitable zone of a solar-like star with a solar-like activity cycle is around 6 Earth masses (when using five observing runs distributed over a full year, and total length of observations of five years).

\cite{Lagrange10} investigated the detectability of the Earth in the habitable zone around a solar-like star. They concluded that with the highest precision instruments one can only detect an Earth mass planet after several years of intensive monitoring, and then preferably during the low activity phase of the star. In another investigation \cite{Dumusque11b} investigate the detection limits with a HARPS-like instrument, taking into account oscillation, granulation, and activity effects. They study different observing strategies to minimise the stellar noise and conclude that applying three measurements per night of 10 minutes every three days, 10 nights a month is the best strategy out of the ones they tested. Depending on the activity level this strategy would allow to detect 2.5--3.5 Earth mass planets in the habitable zone of an early K dwarf. Still, this means a planet larger than the Earth and in a habitable zone that is closer to the star than 1~AU.

\subsection{Future investigations}

The codes developed here are very versatile and large variety of spot configurations and exoplanets on different orbits can easily be studied. This opens a possibility to do statistical studies of the effect of stellar activity on the detection of exoplanets. One can also use the same methods as are used to detect exoplanets in recovering the input planet from our spectra. The input parameters of the planet are know, and therefore this will also allow for determining at which accuracy different planetary parameters can be recovered. In the second part of this series of papers (\citealt{Andersen14}), we apply the methods developed here to M dwarfs. We will also study in detail the effect stellar activity has on the recovered planetary parameters.

\section{Conclusions}

We have developed methods for investigating the radial velocity jitter caused by starspots. The method allows creating many spot configurations with the same spot filling factors and also enables selecting active latitude and longitude ranges. Planetary signatures can easily be added to the spectra and analysed. From the tests and implementation to solar like stars we can draw the following conclusions:

\begin{itemize}
\item As has been seen in the previous studies observations at longer wavelengths decrease the measured radial velocity jitter. The tests carried out in this study show that the decrease of approximately 50\% in the full jitter amplitude is achieved at wavelengths around 7000~{\AA} in comparison to 3700~{\AA}. Between wavelengths 7000~{\AA} and 9000~{\AA} no significant further decrease in the jitter amplitude is observed.
\item The spectral resolution does not affect the jitter amplitude significantly at the generally used resolving powers of 50,000--130,000 and higher. On the other hand, resolution of 20,000 and less decreases the jitter, but also decreases the measurement accuracy.
\item We verify the previous results showing that the full jitter amplitude depends on the stellar rotational velocity, $v\sin i$. The dependence is linear and even though the exact jitter amplitude depends on the spot size, the slope of the correlation does not.
\item The solar-like activity patterns create largely varying amounts of radial velocity jitter. From a spot coverage factor that represents average solar activity, the full jitter amplitude recovered from our simulated data varies approximately between 1~m/s and 12~m/s. The exact value is driven by how concentrated the spots are.
\item The mean full jitter amplitude varies during the solar-like activity cycle between approximately 1~m/s and 9~m/s. 
\item With realistic observing frequency and solar-like cyclic activity a Neptune sized planet on a one year orbit around a solar mass star can be recovered with high significance. The recovery of an Earth mass planet on a similar orbit on the other hand is very challenging.  
\item The starspots do not only create noise in the radial velocity curves, they can also affect the shape of the radial velocity curve in such a way that the determined orbital parameters change. Eccentricity especially can be affected.
\item The spot surface creation and planet orbit implementation software developed in this study allow for statistical studies of effect of spot jitter in exoplanet detection. These issues are addressed in the second paper in this series.
\end{itemize}

\section*{Acknowledgments}
The authors would like to thank the anonymous referee whose comments helped to improve this paper. H.K. acknowledges the support from the European Commission under the Marie Curie IEF Programme in FP7. J.M.A. acknowledges support through an NSF Graduate Research Fellowship and the Nordic Research Opportunity award.

\appendix

\label{lastpage}

\end{document}